\documentclass[aps,prl,twocolumn,showpacs,superscriptaddress,groupedaddress]{revtex4}
\usepackage{graphics}
\usepackage[normalem]{ulem}
\usepackage{amssymb}
\usepackage{color}
\usepackage{graphicx}
\usepackage{amsthm}
\usepackage{paralist}
\usepackage{verbatim}
\usepackage{chemarr}
\usepackage{braket}

\newtheorem{theorem}{Theorem}
\newtheorem{lemma}[theorem]{Lemma}

\def\EQ#1{\begin{eqnarray}#1\end{eqnarray}}
\def\>{\rangle}
\def\<{\langle}

\newcommand{\Tr}{\operatorname{Tr}}

\newcommand{\ketbra}[2]{|#1\rangle\langle #2|}
\newcommand{\map}[1]{\mathcal{#1}}

\begin{document}

\title{Enhanced Communication With the Assistance of Indefinite Causal Order}

\author{Daniel Ebler}
\affiliation{Department of Computer Science
The University of Hong Kong, Pokfulam Road, Hong Kong}
\affiliation{HKU Shenzhen Institute of Research and Innovation, Kejizhong 2nd Road, Shenzhen, China}
\author{Sina Salek}
\affiliation{Department of Computer Science
The University of Hong Kong, Pokfulam Road, Hong Kong}
\author{Giulio Chiribella}
\affiliation{Department of Computer Science, University of Oxford, Wolfson Building, Parks Road, Oxford OX1 3QD, United Kingdom}
\affiliation{Canadian Institute for Advanced Research,
CIFAR Program in Quantum Information Science, Toronto, ON M5G 1Z8}
\affiliation{HKU Shenzhen Institute of Research and Innovation, Kejizhong 2nd Road, Shenzhen, China}

\begin{abstract}
In quantum Shannon theory, the way information is encoded and decoded takes advantage of the laws of quantum mechanics, while the way communication channels are 
interlinked is assumed to be classical. In this Letter we relax the assumption that quantum channels are combined classically, showing that a quantum communication network where quantum channels are combined in a superposition of different orders can achieve tasks that are impossible in conventional quantum Shannon theory. In particular, we show that two identical copies of a completely depolarizing channel become able to transmit
information when they are combined in a quantum superposition of two alternative orders.  This finding runs counter to the intuition that if two communication channels are 
identical, using them in different orders should not make any difference. The failure of such intuition stems from the fact that a single noisy channel can be a random mixture of elementary, 
non-commuting processes, whose order (or lack thereof) can  affect the ability to transmit information.
\end{abstract}

\maketitle

%\pacs{}

\emph{Introduction \label{intro}} -- Information theory, initiated by the seminal work of Claude Shannon, \cite{Shannon1948}, has given us a framework to understand the fundamental workings of communication, data storage and signal processing.
Shannon's theory was originally formulated with the assumption that the carriers of information and the communication channels are classical. The data were represented by  classical random variables and the 
communication channels were treated as stochastic transition matrices.   However, the laws of nature are fundamentally quantum and one can take advantage of these laws to build  a new model of information processing. This gave rise to quantum Shannon theory \cite{Wilde2013},
where quantum features such as superposition and entanglement were used to enhance communication, increasing transmission rates \cite{Bennett1999}, providing unconditional security \cite{BennettCh1984},
and introducing new means of information transmission \cite{Bennett1993}, just to name a few examples. Nevertheless, quantum Shannon theory is still conservative, in  that it assumes that the communication channels are combined in a well-defined configuration.  
In principle, quantum theory allows for new ways to combine communication channels, by connecting them in a quantum superposition of different configurations.  In particular, quantum theory allows the 
order of application of channels to be entangled with a control system \cite{Chiribella2013}, a situation that is sometimes referred to as a {\em quantum superposition of orders}.  Even more generally, quantum theory allows for  exotic configurations that are not compatible with any  underlying  model where the order is definite \cite{Oreshkov2012}. Both features could emerge in a theory of 
quantum gravity \cite{Hardy2005,Hardy2009} and would offer enhancements in a number of tasks, such as testing properties of quantum channels \cite{Chiribella2012a, Araujo2014}, playing non-local games \cite{Oreshkov2012},
and reducing communication complexity \cite{Guerin2016}. 
In this Letter, we show that the ability to combine quantum channels in a superposition of orders can boost the rate of communication beyond the limits of conventional quantum Shannon theory.

\begin{figure}
  \centering%
      \includegraphics[width=0.48\textwidth]{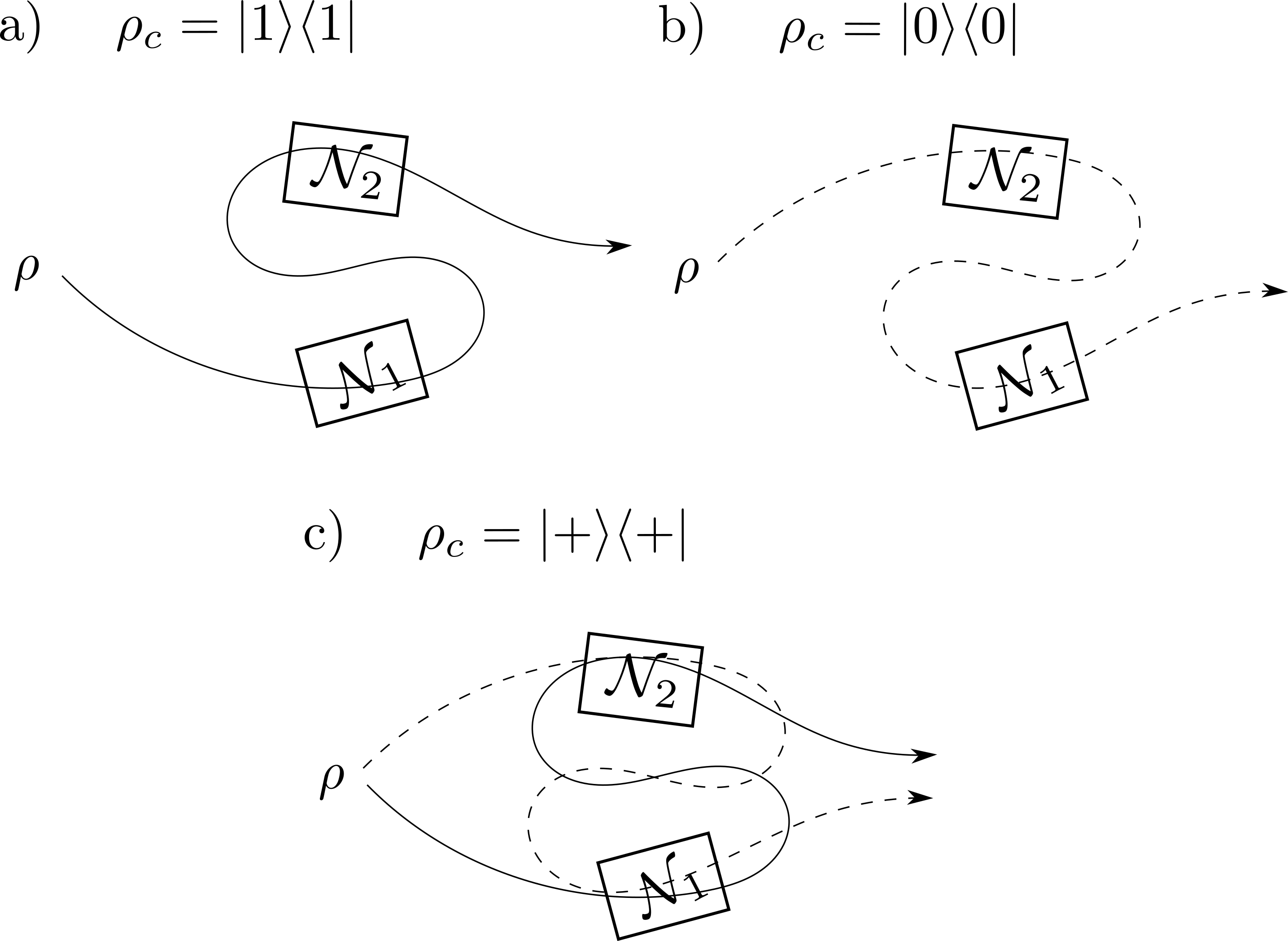}
  \caption{ {\bf Fixed order vs  superposition of orders.}  1(a) A quantum particle, prepared in the state $\rho$, goes first through channel $\map N_1$ and then through channel $\map N_2$.  This configuration is associated to the state $\rho_c =  |1\>\<1|$ of a control qubit, in which the choice of order is encoded.   1(b)  The quantum particle goes first through $\map N_2$ and then through $\map N_1$. This alternative configuration is associated to the qubit state $\rho_c=  |0\>\<0|$.    1(c) The quantum SWITCH creates a superposition of the two configurations 1(a) and 1(b). It takes a control qubit in a superposition state, such as  $\rho_c=\ketbra{+}{+}$, and correlates the order of the two channels with the state of the qubit.  }
    \label{diag}
\end{figure}
 
Our result is based on  a novel quantum primitive,  called the quantum SWITCH \cite{Chiribella2013}. The quantum SWITCH is an operation that takes two channels $\mathcal{N}_1$ and $\mathcal{N}_2$ as inputs
and creates a new channel, which uses the channels $\map N_1$  and $\map N_2$ in an order that is entangled with the state of a control qubit, thus generating a quantum superposition of two alternative orders.    Fig. \ref{diag} illustrates different ways of combining the two channels  $\map N_1$ and $\map N_2$, either in a definite order  or in a quantum superposition of  orders.  In \cite{Chiribella2013} it was shown that the quantum SWITCH cannot be realized by any circuit where the order to the two channels $\map N_1$ and $\map N_2$ is fixed. Likewise, the quantum SWITCH cannot be realized as a classical mixture of circuits using  channels $\map N_1$ and $\map N_2$ in  fixed orders \cite{Chiribella2012a}.  An even broader sense in which the quantum SWITCH cannot be decomposed into quantum processes with definite order has been discussed in \cite{Oreshkov2016}.

In this Letter,  we introduce a quantum Shannon theoretic task where the quantum SWITCH enables two communicating parties to transmit information. We show that  the entanglement of the control system with the order of application of two channels can be used to perform communication
tasks that are impossible when the order is fixed, or even correlated with a classical variable.     Surprisingly, the advantage can be achieved by switching two copies of the same channel, a phenomenon which we refer to as {\em self-switching}.   Specifically, we show 
the advantage of switching two copies of the completely  depolarising channel, which transforms every quantum state into the maximally mixed state. 
Clearly, none of the two fixed configurations  in Fig. (\ref{diag}) can be used to communicate  information.   Here we show that the entanglement of these two configurations with a control system can be used to  communicate classical information -- a phenomenon we call Causal Activation.  This phenomenon sheds light on the fact that the
properties of channels do not solely depend on the way they are 
constructed, but also the way they are {\em combined}: two channels combined in a superposition of different orders behave very differently from  the same channels combined  in a fixed order. 

One of the main pillars of  Shannon theory is quantifying the capacity of channels to communicate information. Channel capacity theorems are of fundamental importance both for the theoretical characterisation of channels, as well as 
the experimental implementation of communication protocols.  In this Letter, we derive an analytical expression for  the Holevo  capacity \cite{Holevo1998}  of the two causally activated depolarizing channels.  Quite counterintuitively, we find that the Holevo capacity is maximum for qubit channels and decreases with the dimension of the input.

\emph{Preliminaries} -- In this section we  review the  concepts and tools needed to understand  our finding. 

We use quantum channels to represent transmission lines in a quantum communication network.  Mathematically, quantum channels are described by completely positive trace preserving maps (CPTP). We will often use the Kraus  decomposition, which allows one to represent the action of a channel $\mathcal{N}$ on a quantum state $\rho$ as $\mathcal{N}(\rho)=\sum_i K_i\rho K_i^\dagger$, where  $\{K_i \}$ is  a set of operators  such 
that $\sum_i K_i^\dagger K_i=I$. 

In a fixed causal structure channels can be composed either in parallel or series. Two channels $\mathcal{N}_1$ and $\mathcal{N}_2$ composed in  parallel  are represented by the tensor product of the two 
channels $\mathcal{N}_1 \otimes \mathcal{N}_2$. If used in series, the second channel simply acts on the output of the first channel, \emph{{{i.e.}}} $\mathcal{N}_2 \circ \mathcal{N}_1$.  More generally, the two 
channels can be connected  in an arbitrary quantum circuit including intermediate operations.

In principle, however, the order does not have to be fixed.  Two channels could be  combined by the quantum SWITCH operation, ending up in a situation where their relative order is  entangled with a control system.  Let us denote the Kraus operators of the channel $\mathcal{N}_1$ as $\{K_i^{(1)}\}$ and $\mathcal{N}_2$ as $\{K_i^{(2)}\}$. The quantum SWITCH uses an auxiliary quantum 
system to control the order of the Kraus operators of the two channels in an indefinite causal manner. The Kraus operators of the overall quantum channel resulting from the switching of $\mathcal{N}_1$ and 
$\mathcal{N}_2$ are
\EQ{
W_{ij}=K_i^{(2)}K_j^{(1)}\otimes \ketbra{0}{0}_{c} + K_j^{(1)}K_i^{(2)}\otimes \ketbra{1}{1}_{c}, \label{SwitchKraus}
}
acting on a target quantum state $\rho$ and a control state $\rho_c$. The action of the quantum SWITCH is then given by
\EQ{
\map S (\map N_1, \map N_2)\left(\rho \otimes \rho_c \right)= \sum_{i,j} W_{ij} \left(\rho \otimes \rho_c \right) W_{ij}^\dagger.  \label{SWITCH}
}
It is easy to check that above definition is independent of the choice of Kraus operators for the channels $\map N_1$ and $\map N_2$. 
Mathematically, the quantum SWITCH   is a {\em higher-order operation} \cite{Chiribella2013}: it takes two channels $\map N_1$ and $\map N_2$ as input and creates a quantum channel $\map S (\map N_1, \map N_2)$ as output. Specifically, this higher-order operation combines the two input channels in an order that depends on the state of the control qubit:  if the qubit is in the state $\rho_x= |0\>\<0|$, the channel $\map S (\map N_1, \map N_2)$ will return the state $\map N_2\map N_1 (\rho) \otimes |0\>\<0|$, if 
the qubit is in the state $\rho_c  = |1\>\<1|$, the channel will return the state $\map N_1\map N_2 (\rho) \otimes |1\>\<1|$. When the qubit is in a superposition of $|0\>$ and $|1\>$, the channel returns a 
correlated state, which can be interpreted as the result of the input channels $\map N_1$ and $\map N_2$ acting on $\rho$ in a quantum superposition of two alternative orders.   

{\em Quantum Shannon theory with the assistance of the quantum SWITCH.}  In quantum Shannon theory, quantum channels represent communication resources. Hence, higher-order operations, like the quantum SWITCH, can be viewed as transformations of resources.  
  Quantum Shannon theory can be cast in the form of a resource theory, by specifying a set of higher-order operations that are regarded as {\em free} \cite{Coecke2016}.
A basic  type of free operation maps an input channel $\map N$ into an output channel $ \map D \circ \map N \circ \map E$,  where $\map E$ and $\map D$ are two  channels,  representing encoding and decoding operations at the sender's and receiver's end, respectively.  Another type is composition in parallel, whereby two  channels $\map N_1$ and $\map N_2$ are combined into the channel $\map N_1\otimes \map N_2$.  Finally,  it is also natural to  consider scenarios where the sender sends the information to the receiver through a repeater, connected to the sender and the receiver through two channels $\map N_1$ and $\map N_2$, respectively.   The corresponding type of free operation is composition in sequence, whereby two input channels  $\map N_1$ and $\map N_2$ are combined into the output channel  $\map N_2  \circ \map R \circ \map N_1$, where  $\map R$ represents the operation performed by the repeater.     Combining these  three types of operations (possibly including free classical correlations), one obtains a resource theory, suitable to describe basic communication tasks involving a single sender and a single receiver.

We now define an extension of standard quantum Shannon theory that includes quantum  superpositions of causal orders.  We do this in the  resource-theoretic framework, by adding the quantum SWITCH to the set of free operations.  More precisely, we add the free operation that maps a pair of channels $\map N_1$ and $\map N_2$ to the new channel $\map N'$ defined by $\map N'(\rho)   =  \map S  (\map N_1,\map N_2)  ( \rho\otimes \rho_c)$, where $\map S$ is defined as in Eq. (\ref{SWITCH}) and $\rho_c$ is a fixed state of the control qubit. Note that the state of the control   is part of the way the two channels are combined, and is
{\em not} accessible to the sender:  the sender cannot encode classical bit in the state of the control. The control is only accessible to the receiver, who can use it as an aid for decoding.   We refer to the extended model as {\em quantum Shannon theory with the assistance of the quantum SWITCH}.  Adding the quantum SWITCH is similar to what is done in other variants of quantum  Shannon theory, where one adds free entanglement \cite{Bennett1999},  free symmetric side-channels \cite{Smith2008b,Smith2008a}, free no-signalling channels  \cite{cubitt2011zero,duan2016no} and such like.

In the following we will focus on the communication of classical information. Holevo \cite{Holevo1998}, Schumacher, and Westmoreland \cite{Schumacher1997}(HSW) proved that a single copy of any 
quantum channel $\map N$ can communicate classical information at best at the rate $\chi(\map N):= \max_{\{p_x,\rho_x\}} I(X;B)_\sigma$, where $ I(X;B)_\sigma$ is the von Neumann Mutual Information, evaluated on a state of the form  $\sigma :=\sum_x p_x \ketbra{x}{x}_X \otimes \map N (\rho_x)_B$ and maximized over all possible ensembles $\{  p_x, \rho_x\}$. The quantity $\chi  (\map N) $  is called the Holevo information and  has been shown to be in general non-additive \cite{Hastings2009}.
This means that there exist two channels $\map N$ and $\map M$ such that $\chi(\map N \otimes \map M) > \chi(\map N) + \chi(\map M)$. Therefore, if many copies of a channel whose Holevo information 
is non-additive are available, the Holevo information  is a lower bound for the capacity of quantum channels to communicate classical information. Operationally, the lower bound corresponds to the amount of information that can be transmitted if the sender uses only  product states in the encoding.

One of the implications of the HSW theorem is that any quantum channel that is not constant can be used to communicate classical information. This is because for a channel $\map N$ that is not constant, there exist at least two 
pure states $\ket{\phi}$ and $\ket{\psi}$, such that $\map N(\ketbra{\phi}{\phi})\ne \map N (\ketbra{\psi}{\psi})$. Using these two states  with equal probability it is immediate to see that the Holevo information is positive.   On the other hand,  the Holevo information  of  a constant channel is trivially zero. Even if the constant  channel is used  many times, none of the operations allowed in the standard model of quantum Shannon theory allows to generate a channel that transmits information.  In the following  we will show that, in contrast, classical communication can become possible  with the assistance of  the quantum SWITCH.

\emph{Main Result} -- A completely depolarising channel $\map N^{\rm D}$ on a $d$-dimensional quantum system can be represented by uniform randomisation over $d^2$ orthogonal unitary operators $U_i$, such that its action on a state $\rho$ is
\EQ{
\map N^{\rm D}(\rho)= \frac{1}{d^2} \sum_{i=1}^{d^2} U_i \rho U_i^\dagger =  \Tr[\rho] \frac{I}{d}. \label{depo}
}

Therefore, according to Eq. (\ref{SwitchKraus}), the overall quantum channel resulting from the quantum SWITCH of two completely depolarising channels has Kraus operators

\EQ{
W_{ij}=\frac{1}{d^2} \left( U_i U_j \otimes  \ketbra{0}{0}_{c} + U_j U_i \otimes \ketbra{1}{1}_{c}\right).\label{krausdoubledep}
}

Suppose that the control system is fixed to the state $\rho_c:=\ketbra{\psi_c}{\psi_c}$, where $\ket{\psi_c}:=\sqrt{p} \ket{0}+\sqrt{1-p}\ket{1}$. If the sender prepares the target system in the state $\rho$, then the receiver will get the output state 
\begin{widetext}
\EQ{
\map S (\map N^D, \map N^D)\left(\rho \otimes \rho_c \right)&=&\frac{1}{d^4} \sum_{i,j} \Big( p\ketbra{0}{0}_{{{c}}} \otimes U_i U_j \rho U_j^\dagger U_i^\dagger  + (1-p) \ketbra{1}{1}_{{{c}}} \otimes U_j U_i \rho U_i^\dagger U_j^\dagger  \nonumber \\
&& + \sqrt{p(1-p)} \ketbra{0}{1}_{{{c}}} \otimes U_i U_j \rho U_i^\dagger U_j^\dagger + \sqrt{p(1-p)} \ketbra{1}{0}_{{{c}}} \otimes U_j U_i \rho U_j^\dagger U_i^\dagger  \Big) \nonumber \\
&=& p \ketbra{0}{0}_{{{c}}} \otimes \frac{I}{d}  + (1-p) \ketbra{1}{1}_{{{c}}} \otimes \frac{I}{d} + \frac{\sqrt{p(1-p)}}{d^2} \ketbra{0}{1}_{{{c}}} \otimes \sum_j \Tr[U_j \rho] \frac{U_j^\dagger}{d}  \nonumber \\
&& + \frac{\sqrt{p(1-p)}}{d^2} \ketbra{1}{0}_{{{c}}} \otimes \sum_j \Tr[\rho U_j^\dagger] \frac{U_j}{d} \nonumber \\
&=&(p\ketbra{0}{0}_{{{c}}} + (1-p) \ketbra{1}{1}_{{{c}}}) \otimes \frac{I}{d} + \sqrt{p(1-p)} (\ketbra{0}{1}_{{{c}}} + \ketbra{1}{0}_{{{c}}}) \otimes \frac{\rho}{d^2} \ .\label{doubledeph}
}
\end{widetext}
The first equality follows from Eq. (\ref{krausdoubledep}).   The second equality is the application of the depolarising channel in Eq. (\ref{depo}). Finally, the last equation follows from the fact that the operators 
$U_j$ form an orthonormal basis for the set of $d\times d$ matrices.

Eq. (\ref{doubledeph}) shows that the quantum SWITCH of two depolarising channels has  a clear dependence on the input state $\rho$.   Therefore, the HSW theorem 
implies that we can communicate classical information at a non-zero rate.

The quantum SWITCH implements a transfer of information from the input system to the correlations between the output system and the control.   Note that  the information is not contained in the state of the system alone, nor in the state of the control alone: it is genuinely contained in the correlations.  Note also that these correlations must be quantum:  if the control decoheres in the basis $\{|0\>,  |1\>\}$, the information is completely lost.   In spite of this,  the receiver does not need to perform entangled operations in the decoding.   Instead, the receiver  measure   the control system in the Fourier basis 
$\lbrace |+\> , |-\> \rbrace$, obtaining the conditional states 
\EQ{
\< \pm |  \map S (\map N^D, \map N^D)\left(\rho \otimes \rho_c \right) |\pm\> = \frac{I}{2d} \pm \sqrt{p(1-p)} \frac{\rho}{d^2} \ .
}
Since these states still depend on $\rho$, the receiver can use this dependence to extract non-zero information from the target system. In terms of the Kraus operators (\ref{krausdoubledep}), the postselection on 
the outcomes $+$ and $-$ generates the noisy channel generalisation of the {\em quantum superpositions of time evolutions}  proposed by Aharonov and collaborators   \cite{Aharonov1990}. 

So far we have shown that the quantum SWITCH allows one to use depolarising channels to communicate at {\em some} non-zero rate.  We now compute the optimal rate in the case of product encodings, by analytically calculating the maximum Holevo information over all input ensembles. We restrict our
attention to the case where
the control qubit is in  the state $\rho_c=\ketbra{+}{+}$, because for such state the communication rate is the highest. The expression for
the maximum  Holevo information is derived in the Supplemental Material, where, in fact, we derive an even more general expression, valid for arbitrary depolarizing channels, 
sending an input  state $\rho$
to an output state $q\rho+(1-q)I/d$, with $0\leq q \leq 1$ a generic noise parameter. For the complete depolarizing channel ($q=0$), we find the Holevo information  to be
\begin{widetext}
\EQ{
\chi(S (\map N^D, \map N^D))&=&\log d +H(\widetilde{\rho_c}) \nonumber\\
&&+ \left\lbrace  \left( \frac{d+1}{2d^2} \right) \log \left( \frac{d+1}{2d^2} \right) + \left( \frac{d-1}{2d^2} \right) \log \left( \frac{d-1}{2d^2} \right) + 2(d-1) \left( \frac{1}{2d} \right) \log \left( \frac{1}{2d} \right) \right\rbrace, \label{chiMain}
}
\end{widetext}
where $\widetilde{\rho_c}= 1/2 \ketbra{0}{0}  + 1/2\ketbra{1}{1}+ 1/2d^2 (\ketbra{0}{1}+\ketbra{1}{0})$ is the reduced state of the control system.
It should not be surprising that the entropy of the control system appears in the expression for the capacity of the switched depolarising channels. This is because the control system is a parameter describing the  way in which the depolarising channels are combined. 

Eq. (\ref{chiMain}) is the best rate one can communicate classical information by switching only two copies of depolarising channel.
However, if one has access to more copies of these channels, one may be able communicate more by inputting states that are entangled across the channels. To show this would require a proof that the overall mapping 
generated by switching depolarising channels is not additive. Since this question is separate from the main point of this Letter, we leave this task for future investigation. 
  
\emph{Conclusions and Discussion} -- In this work, we explored an extension of quantum Shannon theory where communication channels can be combined in a quantum superposition of orders. 
In this extended model,  we showed that combining two completely depolarizing channels with the quantum SWITCH activates them, allowing the transmission of classical information. In contrast, no such activation is possible in the standard model, %no matter how many times the depolarizing channels are used. 
 where the order is fixed  or controlled by a classical random variable.  

Strikingly, we showed that the Shannon theoretic advantage can be gained as a result of creating a superposition of a
channel with another copy of itself. This result  may seem paradoxical, because  exchanging two uses of the same channel would not have any observable effect in any ordinary quantum circuit. The resolution of the paradox lies in the fact that noisy quantum channels can be seen as random mixture of different processes, corresponding to different Kraus operators. The advantage of the self-switching arises because some of these  processes do not commute with each other,  and therefore a quantum control on the order offers a non-trivial resource.   We  observe that no self-switching effect arises for quantum channels that admit a Kraus decomposition consisting of mutually commuting operators. 

Our results are an invitation to investigate a new paradigm of Shannon theory, where the order of the communication channels can be in a quantum superposition.  This paradigm may find applications in future quantum communication networks. Consider a situation where a provider connects different communicating parties through a network of quantum channels. In this situation, the provider could opt 
to connect the channels in a superposition of alternative configurations, thereby  boosting the communication rates between parties.  
Of course, every such  application requires a careful analysis of  physical resources required for the implementation of the quantum SWITCH. While  in this work we treated the quantum SWITCH as an abstract higher-order operation, there exist different ways in which this operation could be realized, including table-top photonic implementations  \cite{Procopio2015,Rubino2017},
implementations with ion traps \cite{friis2014implementing}  and superconducting circuits \cite{friis2015coherent}.   The practical extent of the advantage shown in our paper greatly depends on the resources required in each implementation. For example, Oreshkov \cite{Oreshkov2018} has recently analyzed the structure of the photonic implementations of  \cite{Procopio2015,Rubino2017}, showing that an essential ingredient is the ability to delocalize the input channels in time, coherently controlling when the environment interacts with the system. 
 On the other hand, our result provides new motivation to the development of experimental techniques for   the implementation of the quantum SWITCH.

\begin{acknowledgments}
We thank Philippe Grangier, Renato Renner, Aephraim Steinberg, Ognyan Oreshkov, Philip Walter, Giulia Rubino, Paolo Perinotti, Michal Sedl\'{a}k, Matt Leifer and Christa Zoufal, for helpful and engaging discussions 
during the ``Hong Kong Workshop on Quantum Information and Foundations 2018".   This work is supported by the National
Natural Science Foundation of China through grant 11675136, the Croucher Foundation, the Canadian Institute for Advanced
Research (CIFAR), the  Hong Research Grant Council through grant 17326616, and the Foundational
Questions Institute through grant FQXi-RFP3-1325.  This publication was
made possible through the support of a grant from the John Templeton Foundation. The opinions expressed in this
publication are those of the authors and do not necessarily reflect the views of the John Templeton Foundation. 
 \end{acknowledgments}
 
 {\small
\bibliography{SupAct.bib}
\bibliographystyle{plain}
}
\appendix
\onecolumngrid

\section{Lower bound for the classical capacity of the Quantum SWITCH of two depolarizing channels}

In this supplemental material we give the expression for the Holevo information of two depolarising channels, used in a Quantum SWITCH with a control qubit initially in the state $|+\>$. In order to do this, we first show how such an operation acts on a general quantum state.

A generic depolarising channel can be represented as 
\EQ{
\map N_q^D(\rho)
&=&q \cdot \rho + (1-q) \cdot \Tr[\rho] \frac{I}{d} \\
&=& q \cdot \rho + \frac{1-q}{d^2} \sum_{i=1}^{d^2} U_i \rho U_i^\dagger \ ,  \label{pdepo}
}
where $\{ U_i\}_{i=1}^{d^2}$ are unitary operators and form an orthonormal  basis of the space of $d\times d$ matrices. 

When two depolarizing channels are inserted into the Quantum SWITCH, one obtains a quantum channel $\map S  (\map N_q^D, \map N_q^D)$, whose Kraus operators are given by Eq. (4). Introducing the notation  $U_0:=\frac{d \sqrt{q}}{\sqrt{1-q}} \cdot I$, we can express the Kraus operators as 
\EQ{
W_{ij}=\frac{1-q}{d^2} \left( U_i U_j \otimes  \ketbra{0}{0}_{c} + U_j U_i \otimes \ketbra{1}{1}_{c}\right),  \qquad \forall  i, j \in   \{0,1,\dots,  d^2\} \, .
}
Let the control system be in the pure state $\rho_c=p\ketbra{0}{0}+(1-p)\ketbra{1}{1}+\sqrt{p(1-p)}(\ketbra{0}{1}+\ketbra{1}{0})$. Using this control system, the action of the new channel  $\map S  (\map N_q^D, \map N_q^D)$   on a  generic state
$\rho$ is 
\EQ{
\map S (\map N_q^D, \map N_q^D)\left(\rho \otimes \rho_c \right)&=&\frac{(1-q)^2}{d^4} \sum_{i,j = 1}^{d^2} \Big( p\ketbra{0}{0} \otimes U_i U_j \rho U_j^\dagger U_i^\dagger  + (1-p) \ketbra{1}{1} \otimes U_j U_i \rho U_i^\dagger U_j^\dagger  \nonumber \\
&&+ \sqrt{p(1-p)} \ketbra{0}{1} \otimes U_i U_j \rho U_i^\dagger U_j^\dagger + \sqrt{p(1-p)} \ketbra{1}{0} \otimes U_j U_i \rho U_j^\dagger U_i^\dagger  \Big) \nonumber \\
&&+ \frac{q(1-q)}{d^2} \sum_{i=1}^{d^2} \Big( p\ketbra{0}{0} \otimes U_i \rho U_i^\dagger  + (1-p) \ketbra{1}{1} \otimes U_i \rho U_i^\dagger \nonumber \\
&&+ \sqrt{p(1-p)} \ketbra{0}{1} \otimes U_i \rho U_i^\dagger + \sqrt{p(1-p)} \ketbra{1}{0} \otimes  U_i \rho U_i^\dagger  \Big) \nonumber \\
&&+ \frac{q(1-q)}{d^2} \sum_{j=1}^{d^2} \Big( p\ketbra{0}{0} \otimes U_j \rho U_j^\dagger  + (1-p) \ketbra{1}{1} \otimes U_j \rho U_j^\dagger \nonumber \\
&&+ \sqrt{p(1-p)} \ketbra{0}{1} \otimes U_j \rho U_j^\dagger + \sqrt{p(1-p)} \ketbra{1}{0} \otimes  U_j \rho U_j^\dagger  \Big) \nonumber \\
&&+ q^2 \cdot \rho \otimes \rho_c \ .\label{ddoubledeph}
}
The four contributions arise in the following way: first, we fix both $i \neq 0$ and $j \neq 0$. Second, we fix $j=0$ and $i \neq 0$. Third, we swap the roles to $i=0$ and $j \neq 0$. Finally, both $i=0$ and $j=0$
gives the last term. Since the randomization over the set of unitaries $U_{j \neq 0}$ completely depolarises the state, i.~e.~$\frac{1}{d^2}\sum_{i} U_i \rho U_i^\dagger =\frac{I}{d}$,
the only non-trivial contribution is the first part for which $i \neq 0$ and $j \neq 0$. However, this term is the same as Eq. (5) in the main manuscript.
This gives the following output of the SWITCH with two depolarising channels
\EQ{
\map S (\map N_q^D, \map N_q^D)\left(\rho \otimes \rho_c \right)&=& (1-q)^2 \left(  ( p\ketbra{0}{0}_{{{c}}} + (1-p) \ketbra{1}{1}_{{{c}}}) \otimes \frac{I}{d} + \sqrt{p(1-p)} (\ketbra{0}{1}_{{{c}}} + \ketbra{1}{0}_{{{c}}}) \otimes \frac{\rho}{d^2}   \right) \nonumber \\
&&+ 2q(1-q) \left( \rho_c \otimes \frac{I}{d} \right) + q^2 \rho_c \otimes \rho \ . \label{pdoubledeph}
}

We can now proceed to computing the best rate at which this channel can communicate classical information.
 To this purpose, we compute the quantum mutual information of  the quantum-classical state.
\EQ{
\sigma_{XBC}:=\sum_x p_x \ketbra{x}{x}_X \otimes  \map S (\map N_q^D, \map N_q^D)  (\rho_a^x \otimes \rho_c) \ ,
}
where the lower scripts denote explicitly the Hilbert spaces of the state. It has been shown that it is sufficient to optimise Holevo information over classical-quantum states with conditional states that are pure 
(see \emph{e.g.} Theorem 13.3.2 of \cite{Wilde2013}). The Holevo information is a lower bound of the classical capacity of a channel. Hence,  in order to bound the classical capacity of the channel 
$\map S (\map N_q^D, \map N_q^D) $, we find the upper limit of the Holevo information. Therefore, we 
define the following extended input state with pure conditional state
\EQ{
\omega_{XIJAC}=\frac{1}{d^2} \sum_{x,i,j} p_x |x\>\<x|_X \otimes |i\>\<i|_I \otimes |j\>\<j|_J \otimes  X(i) Z(j) \Psi^x_A Z(j)^\dagger X(i)^\dagger \otimes \Phi_c \ ,
}
where system $I$ and $J$ are two new classical registers and $\Psi_A^x$ and $\Phi_c$ are some pure states of the target and control systems respectively. Besides, the probability distribution for the
registers $I$ and $J$ was assumed to be uniform, leading to the pre-factor $1/d^2$. Here, $X(i)$ and $Z(j)$ denote the generalized Pauli operators acting on a vector $|l\>$ as
\EQ{
X(i)|l\>&=&|i \oplus l\> \nonumber \\
Z(j)|l\>&=&e^{2 \pi i j l /d} |l\> \ .
}

The mutual information $I(X;BC)_{\sigma}$   can be bounded as  
\EQ{
I(X;BC)_\sigma&=&H(BC)_\sigma-H(BC|X)_\sigma \nonumber \\
&\leq&H(BC)_\omega-H(BC|X)_\sigma \nonumber \\
&= & \log d + H(\widetilde{\rho_c}) - H(BC|X)_\sigma \ . \label{bound1}
}
The first inequality follows from concavity of the von Neumann entropy. The last line is a direct consequence of   
\EQ{
\Tr_{XIJ}[(\map S (\map N_q^D, \map N_q^D) \otimes I)(\omega_{XIJAC})] &=& \frac{I_{A}}{d} \otimes \widetilde{\rho_c} \ ,
}
with 
\EQ{
\widetilde{\rho_c}= (1-q)^2 \left( p \ketbra{0}{0}_{{{c}}}  + (1-p)\ketbra{1}{1}_{{{c}}}  + \sqrt{p(1-p)}/d^2  (\ketbra{0}{1}_{{{c}}}+ \ketbra{1}{0}_{{{c}}}) \right) + q(2-q) \rho_c \ .
}
above, we assumed that $\Phi_c$ is a general pure state on the control. In Eq. (\ref{bound1}), we then have $\widetilde{\rho_c}$ instead, but replaced by concavity $\sigma$ with $\omega$ before. 
To further upper bound the mutual information, we analyse the conditional entropy $H(BC|X)_\sigma$. 
Define $\theta_{x,i,j}:=S (\map N^D, \map N^D)(X(i)Z(j) \Psi_A^x Z(j)^\dagger X(i)^\dagger \otimes \Phi_c)$
and $\gamma_{x,i,j}:=X(i)Z(j) \map S (\map N^D, \map N^D)( \Psi_A^x  \otimes \Phi_c)Z(j)^\dagger X(i)^\dagger$. We find that
\EQ{
H(BC|XIJ)_\omega&=& \frac{1}{d^2} \sum_{x,i,j} p_x H(BC)_{\theta_{x,i,j}} \nonumber \\
&\stackrel{(\star)}=& \frac{1}{d^2} \sum_{x,i,j} p_x H(BC)_{\gamma_{x,i,j}} \nonumber \\
&=& \sum_{x} p_x H(BC)_{ S (\map N^D, \map N^D)( \Psi_A^x  \otimes \Phi_c)} \nonumber \\
&=& H(BC|X)_\sigma .
}
The second line follows directly from Eq.~(\ref{pdoubledeph}): since the operators $X(i)$ and $Z(j)$ only act on the system state and leave the control invariant, the states $\theta_{x,i,j}$ and $\gamma_{x,i,j}$ coincide. The third 
line is a consequence of the von Neumann entropy being invariant under isometric transformations. Using the chain of equalities above, we can further upper bound Eq.~(\ref{bound1}) as
\EQ{
{I(X;BC)}_\sigma&\leq& \log d  + H(\widetilde{\rho_c}) - H(BC|XIJ)_\omega \nonumber \\
&=& \log d  + H(\widetilde{\rho_c}) - \sum_{x} p_x H(B)_{ \map S (\map N^D, \map N^D)( \Psi_x  \otimes \Phi_c)} \nonumber \\ 
&\leq & \log d  + H(\widetilde{\rho_c})- \min_x H(B)_{ \map S (\map N^D, \map N^D)( \Psi_x  \otimes \Phi_c)} \nonumber \\
&\leq & \log d  + H(\widetilde{\rho_c})- H^{\min}(\map S (\map N^D, \map N^D)). \label{bound2}
}
The first inequality follows because the expectation value can never be smaller than the minimum value. In the second inequality, we defined the quantity $H^{\min}(\map N)= \min_\varphi H(\map N(\varphi))$ for a channel $\map N$ and input state $\varphi$.

In the remainder of this section we compute $H^{\min}(S (\map N^D, \map N^D))$. We start from Eq.~(\ref{ddoubledeph}) and denote the right hand side with the matrix
\EQ{
\map S (\map N^D, \map N^D)(\rho\otimes \rho_c)=\begin{pmatrix} A & B \\ B & \widetilde{A} \end{pmatrix}
}
with 
\EQ{
A&=&((1-q)^2 + 2q(1-q) )p \cdot I/d + q^2 p \cdot \rho \nonumber \\
\widetilde{A}&=&((1-q)^2 + 2q(1-q) )(1-p) \cdot I/d + q^2 (1-p) \cdot \rho \nonumber \\
B&=&(1-q)^2 \sqrt{p(1-p)} \cdot \rho/d^2 + 2q(1-q) \sqrt{p(1-p)} \cdot I/d + q^2 p \rho \ .
}
In the following we set $p=1/2$ and retrieve $\rho_c=|+\>\<+|$ from the main text.
This leads to
$A=\widetilde{A}$, such that the resulting matrix is of the form $\begin{pmatrix} A & B \\ B & A \end{pmatrix}$. In order to find the eigenvalues of $S (\map N^D, \map N^D)(\rho\otimes \rho_c)_{p=1/2}$, we use
the following Lemma. 

\begin{lemma}
Consider a $2d \times 2d$ matrix $M=\begin{pmatrix} A & B \\ B & A \end{pmatrix}$ acting on the vector space $V=\mathbb{C}^d \oplus \mathbb{C}^d$. Let $V=W_1 \oplus W_2$, with $W_1=\lbrace( v,v): v\in \mathbb{C}^d \rbrace$ and $W_2=\lbrace( v,-v): v\in \mathbb{C}^d \rbrace$. If $W_1$ and $W_2$ are invariant under the action of $M$, then the eigenvalues of $M$ are the union of the eigenvalues of $A+B$ and $A-B$.
\end{lemma}

Therefore, finding the eigenvalues of the operators $A+B$ and $A-B$ is sufficient, which reads for the case $p=1/2$ 
\EQ{
A + B &=& ((1-q)^2 + 2q(1-q) )p \cdot I/d + q^2 p \cdot \rho +  (1-q)^2 \sqrt{p(1-p)} \cdot \rho/d^2 + 2q(1-q) \sqrt{p(1-p)} \cdot I/d + q^2 p \rho  \nonumber \\
&=& \left( \frac{(1-q)^2}{2} + 2q(1-q) \right) \frac{I}{d} + \left( q^2 + \frac{(1-q)^2}{2d^2} \right) \rho \nonumber \\
\nonumber \\
A - B &=&  ((1-q)^2 + 2q(1-q) )p \cdot I/d + q^2 p \cdot \rho -  (1-q)^2 \sqrt{p(1-p)} \cdot \rho/d^2 - 2q(1-q) \sqrt{p(1-p)} \cdot I/d  -  q^2 p \rho\nonumber \\
&=& \left( \frac{(1-q)^2}{2} \right) \frac{I}{d} - \left( \frac{(1-q)^2}{2d^2} \right) \rho
}
Since $I$ and $\rho$ commute, there exists an invertible operator $T$ such that
\EQ{
A&=& T \ {\rm diag} (\lambda^A_1,\lambda^A_2 , \dots , \lambda^A_d)\ T^{-1} \nonumber \\
B&=& T \ {\rm diag} (\lambda^B_1,\lambda^B_2 , \dots , \lambda^B_d)\ T^{-1} \ ,
}
where $spec(A)=\lbrace \lambda^A_i \rbrace_{i=1}^d$ and $spec(B)=\lbrace \lambda^B_j \rbrace_{j=1}^d$ denote the eigenvalues of $A$ and $B$ respectively. It follows that
\EQ{
(A \pm B)= T \big({\rm diag} (\lambda^A_1,\lambda^A_2 , \dots , \lambda^A_d)\pm {\rm diag} (\lambda^B_1,\lambda^B_2 , \dots , \lambda^B_d)\big) T^{-1}}
such that the eigenvalues of $A \pm B$ read $\lbrace \lambda^A_i \pm \lambda^B_i \rbrace_{j=1}^d$. Since $A$ and $B$ only contain the identity matrix and $\rho$, we do not need to worry about the ordering of the 
eigenvalues, which is determined by the choice of $T$. Hence, the spectra are given by
\EQ{
spec(S (\map N^D, \map N^D)(\rho\otimes \rho_c)_{p=1/2})&=& spec(A + B) \cup spec(A - B) \nonumber \\
&=& \left\lbrace \lambda_i^+ \right\rbrace_{i=1}^d \cup \left\lbrace \lambda_i^- \right\rbrace_{i=1}^d ,
}
where the eigenvalues $\lbrace\lambda_i^+ \rbrace_{i=1}^d$ and $\lbrace\lambda_i^- \rbrace_{i=1}^d$ read
\EQ{
\left\lbrace\lambda_i^+ \right\rbrace_{i=1}^d &=& \left\lbrace  \frac{(1-q)^2 +4q(1-q)}{2d} + \left( q^2 + \frac{(1-q)^2}{2d^2}  \right) \lambda_i^\rho  \right\rbrace_{i=1}^d \nonumber \\
\left\lbrace\lambda_i^- \right\rbrace_{i=1}^d &=& \left\lbrace  \frac{(1-q)^2}{2d^2} (d-\lambda_i^\rho)  \right\rbrace_{i=1}^d \ .
}

Above, we denoted the eigenvalues of the system input as $spec(\rho):=\lbrace \lambda_i^\rho \rbrace_{i=1}^d$. Finally, this yields for the minimum entropy of the channel $S (\map N^D, \map N^D)$
\EQ{
H^{\min}(\map S (\map N^D, \map N^D)  (\rho\otimes \rho_c))&=&\min_\rho H(\map S (\map N^D, \map N^D)  (\rho\otimes \rho_c))\nonumber \\
&=& \min_\rho \sum_i - \left\lbrace  \lambda_i^+ \log \lambda_i^+ + \lambda_i^- \log \lambda_i^- \right\rbrace \ . \label{hmin}
}
Since the entropy is a concave function of the eigenvalues $\lambda^{\rho}_i$, $H^{\min}(S (\map N^D, \map N^D)(\rho\otimes \rho_c))$ only possesses global maxima. Furthermore, the minimum value is attained at the border of the interval 
$[0,1]^{\times d}$. With the restriction that the eigenvalues have to sum up to one, the only valid points on the edges are
\EQ{
\lambda^{\rho}_i&=&1 \nonumber \\
 \lambda^{\rho}_{j \neq i}&=&0 \quad \forall \ j \neq i \ .
}
We note that all $d$ of those points are equivalent. This shows that the optimal $\rho$ that minimizes $H(S (\map N^D, \map N^D)  (\rho\otimes \rho_c))$ is a pure state. Therefore, we finally obtain
\EQ{
H^{\min}(S (\map N^D, \map N^D))&=& - \Bigg\lbrace \left(  \frac{d+1+q(d-1)(2+q(2d-1))}{2d^2} \right) \log\left(\frac{d+1+q(d-1)(2+q(2d-1))}{2d^2} \right) \nonumber \\
&+& (d-1) \left(\frac{(1-q)^2 + 4q(1-q)}{2d} \right) \log\left( \frac{(1-q)^2 + 4q(1-q)}{2d} \right) \nonumber \\
&+&  \left( \frac{(1-q)^2}{2d^2} (d-1) \right) \log\left( \frac{(1-q)^2}{2d^2} (d-1) \right) + (d-1) \left(\frac{(1-q)^2}{2d}\right) \log\left( \frac{(1-q)^2}{2d} \right) \Bigg\rbrace \ .
}
Therefore, we find the following upper bound for the mutual information
\EQ{
I(X;B) \leq \log d +H(\widetilde{\rho_c})- H^{\min}(S (\map N^D, \map N^D)) \ . \label{bound3}
}
It is left to find the state ensemble $\lbrace p_x, \rho_x \rbrace$ that achieves this bound. We directly find that picking the family $\lbrace \rho_x \rbrace$ as $d$ pure orthonormal states indeed satisfies 
Eq.~(\ref{bound3}) with equality. Altogether, the state $\omega_{XA}= 1/d \sum_i |i\>\<i| \otimes |i\>\< i|$ is optimal. 
We find the chi-quantity as a function of the depolarising parameter $q$ to be
\EQ{
\chi(S (\map N^D, \map N^D))&=& \max_\rho I(X;B) \nonumber \\
&=& \log d  + H(\widetilde{\rho_c}) - H^{\min}(S (\map N^D, \map N^D)) \ .
}
\end{document}